\documentclass{article}
\usepackage{epsfig,graphicx,times,amsmath,amsthm, amssymb, multicol}
\usepackage[T1]{fontenc}  
\usepackage{calc}
\usepackage{enumerate}
\usepackage{subfig}
\usepackage{amsfonts}
\usepackage{amssymb}
\usepackage{multirow}
\usepackage{epstopdf}
\usepackage{cellspace}

\makeatletter
\renewcommand*{\@biblabel}[1]{\hfill#1.}
\makeatother

\begin{document}

\title{I. Determination of chemical reaction rate constants by numerical nonlinear analysis: differential methods }
\author{Christopher G. Jesudason 
\thanks{Emails: jesu@um.edu.my or  chris\_guna@yahoo.com  }\\
{\normalsize   Department of  Chemistry and Center for Theoretical and Computational Physics} \\ 
 {\normalsize University of Malaya}\\
  {\normalsize 50603 Kuala Lumpur, Malaysia} }
 \date{\today}
\maketitle


\begin{abstract} The primary emphasis of this work on kinetics is to illustrate the \emph{a posteriori} approach to applications, where focus on data leads to novel outcomes, rather than the \emph{a priori} tendencies of applied analysis which imposes constructs on the nature of the observable. The secondary intention is the development of appropriate methods consonant  with experimental definitions.  Chemical kinetic equations were largely developed with the assumption  of    rate constant invariance  and in particular  rate constant determination   usually required  knowledge of  the  initial concentrations. These   methods could not determine  the instantaneous rate constant. Previous work   based on precise simulation data [ J. Math . Chem {\bf 43} (2008) 976--1023] of a bidirectional chemical reaction  system in equilibrium concluded that the rate constants is a function of species concentration through the defined  and determined reactivity coefficients  for  at least elementary reactions. Inhomogeneities in the reaction medium might also lead to cross-coupling of forces and fluxes, leading to concentration dependencies.    By focusing on  gradients,  it is possible to determine both the average and   instantaneous rate constants that can monitor changes in the rate constant with concentration changes as suggested by this theory. Here,  methods are developed and discussed utilizing nonlinear analysis which does not require  exact knowledge of initial concentrations. These methods are  compared with those derived from  standard methodology for 	known chemical reactions studied by eminent kineticists and  in one case with a reaction whose initial reactant concentration was not well determined. These gradient methods are shown to be consistent with the ones from standard methods and could readily serve as alternatives for studies where there are limits or unknowns in the initial conditions, such as in the burgeoning fields of astrophysics and astrochemistry, forensics, archeology and biology . All four reactions studied exhibited semi sinusoidal-like change with reactant concentration change which standard  methods cannot detect, which seems to constitute the  observation of a  new effect that  is not predicted by current formulations,  where the possibility  that  the observations are due to artifacts from  instrumental  errors or the optimization method is reasoned as unlikely since the experiments were conducted by different groups at very different times with different classes of reactions. Two broad reasons are given for this observation, and experiments are suggested that can  discriminate between these two effects.  Although first  and second order reactions were investigated here using data from prominent experimentalists, the method  applies to arbitrary fractional orders  by polynomial expansion of the rate decay curves where closed form integrated expressions do not exist.   Integral methods for the above will be investigated next.
\end{abstract}
\section{INTRODUCTION AND METHODS} 
Current trends in mathematical applications almost always indicate the creation  of mathematical structures that are then supposed and considered  to mirror physical reality  and experimental outcomes. Less common  are applications that analyze  experimental data using as closely as possible  the operational definition   of variables to elucidate  the validity or otherwise of theories. The main thrust of this sequel is to illustrate research that puts priority on the experimental data as a means to constructing  or suggesting theoretical  and mathematical structures. The data  from highly empirical field of chemical kinetics is used within the scope of  the definition of the measured variables to discover/uncover  new effects;  it is suggested here based on the outcome of the analysis that mathematical  analysis  of the data in other fields  within the operational definition of the empirical variables  can elucidate new phenomena and suggest how appropriate and consistent  theory  can be constructed \emph{a posteriori}, rather than the \emph{a priori} tendencies of many applied mathematicians.  One aspect of this culture is the cult of  prediction and of predictability in the natural sciences where resources are expended in performing experiments to verify and substantiate theories.  The methods developed here are of secondary importance compared to the \emph{a posteriori} analysis of the data and its outcomes; these methods refer to  variables which come from  the  experimental definition. 
For an elementary  reaction 
\begin{equation} \label{eq:14}
\mbox{A}_1 + \nu_2\mbox{A}_2 \ldots \nu_{n_r}\mbox{A}_{n_r} \rightarrow \mbox{Products}
\end{equation}
we define the rate constant   $k$ as  the factor in the equation
\begin{equation} \label{eq:15}
\frac{d{[\mbox{A}_1]}}{dt}= \dot{\mbox{[A]}}=-k\mbox{[Q]}
\end{equation}
  where
	\[ [Q]=\prod_{i=1}^{n_r} [A_i]^{\nu_{i}}=l_{A_1A_2\ldots A_{n_r}}(t)=l_Q(t)\]
with $l_R(t)=\mbox{[R]}(t)$ in general and $l_A(t)=\mbox{[A]}_1(t)$ in particular, with the notation $l_{A_1}(t)l_{A_2}(t)\ldots l_{A_{n_r}}(t)=l_Q(t)$. The square brackets denote the concentration of the species, and $t$ is the time parameter.  For the above, the order $O$, which need not be integer is defined as $O=\sum_{i=1}^{n_r}\nu_i$. Clearly, for the above
\begin{equation}\label{eq:15x}
	k=-\left\{\frac{d[A_1]}{dt}\right\}/Q.
\end{equation}
We determine $k$ here directly by various methods of computing  average  and instantaneous gradients for equation (\ref{eq:15}). In traditional methods, the integrated rate law expression is known for only a handful of integer $O$ values of (\ref{eq:15}) which also require initial conditions; no such restrictions apply to the current numerical technique.  Another  class of method is through a least squares optimization of the function $R(k)$ for $n$ datapoints defined as 
\begin{equation} \label{eq:15b}
R(k)=\sum_{i=1}^n \left(\frac{dl_A(t_i)}{dt} - kl_Q(t_i) ) \right)^2.
\end{equation}
Then,
\[R^\prime(k)=0\Rightarrow \sum_{i=1}^n \left(\frac{dl_A(t_i)}{dt} - kl_Q(t_i)\right) l_Q(t_i)=0 \]
implies
\begin{equation} \label{eq:16}
k=-\frac{\sum_{i=1}^n \frac{dl_A(t_i)}{dt}.l_Q(t_i)}{\sum^n_{i=1}{l^2_Q(t_i)}}.
\end{equation}
Eq.(\ref{eq:16}) does not require iterative methods such as   Newton-Raphson's (NR) to determine  the rate constant. A variant of the $R(k)$ optimization above is found in Sec(\ref{ss2}); the reason being that we optimize over an intergrated expression rather than directly the rate equation (\ref{eq:15}) such as (\ref{eq:t1}) for the first order rate constant $k_1$and {\ref{eq:t2}} for the second order constant $k_2$. All variants of the above methods will be discussed in sequence in what follows.  
Most kinetic determinations use logarithmic plots with known initial concentrations, although there have been attempts at integral methods \cite[and refs. therein]{moore1,gug1,john1,swin2,swin1,will1,khan1,holt1,went2,went1,houser1, cort1} that dispense with the  initial concentration. (There may be  ambiguities  in  e.g. \cite{moore1} concerning  choice of independent variables that will  be discussed elsewhere). These standard methods all assume constancy of the rate constant $k$, and therefore have not inspired  methodology that can detect the changes to the rate constant that, according to the detailed results of ref.\cite{cgj1} sheds important information on the activation energy profile changes due to the force fields acting on the reacting species. There are conceivably  many other reasons for variation  with time of the rate constants; they include coupling of inhomogenous temperature field gradients  with chemical species fluxes, leading to physical variable inhomogeneities in the reaction cell  that modifies the rate of reaction with time. This is discussed after the data is presented in what follows.

There have been detailed and specialized  reports and treatises of  computational techniques over the many decades but these have been sparse and far between. Wiberg has \cite[p.757]{wib1} described various more advanced series expansion techniques in conjunction  with least squares analysis to derive kinetic data. His use of numerical integration is confined to solving by Runge-Kutta integration a set of coupled equations, such as feature in an enzyme-catalysed reaction \cite[p.771]{wib1}. Wiberg  in turn draws upon the collective efforts  collated by D. F. Detar \cite{detar1,detar2}. It seems that Detar's collation anticipates to some degree many of the developments cited  above in this work's bibliography. A first order treatment of a chemical reaction is given in the program LSKIN1 \cite[p.126]{detar1} requires data and time intervals that are conformable to the Roseveare-Guggenheim time interval requirement. LSKIN2 \cite[p.3]{detar2} solves for rate constant and initial concentrations of a second order reaction based on a series expansion of the integrated rate law expression. Here, the curvature would introduce "errors" if a linear expansion were used. For both these methods, the constancy of the rate constant $k$ is a basic assumption, which is not the case here.

Nonlinear analysis (NLA),  will be attempted here in preliminary form, in order to compute both the instantaneous and average rate constants. We analyze 2 first order reactions and one second order one using  data from prominent kineticists. In addition, we select one first order reaction whose initial concentration index is ambiguous utilizing  the others as a reference  to gauge the likelihood of our result based on NLA; if our analysis concurs with the 3 reactions from the literature, then one might  be confident that  the NLA analysis of the ambiguous reaction is reasonably accurate.  Important experiments in science  are conducted under uncontrolled conditions, such as in astrochemical reaction rate determinations and photochemical emission spectra in the Mars and  Titan atmospheres  over the several decades \cite{astro1,astro2,astro3}. A similar situation obtains in forensic science and archeology and in biological physiological rate determinations. The basic methods presented here caters for both controlled and uncontrolled initial conditions.  

The 3 first order reactions  (i)-(iii) and   second order reaction (iv) studied are itemized below:
\begin{enumerate}[(i)]
\item  the tert butyl chloride  hydrolysis reaction in ethanol solvent (80\%v/v) at $25^o$C  derived  from the Year III teaching laboratory of this University (UM) where the initial concentration, although determined, is ambiguous. Because of time constraints, the inaccurate  $\lambda_\infty=2050 \mu\text{S cm}^{-1}$ for (i) was determined by heating the reaction vessel at the end of the monitoring to $60^o$C until there was no apparent change in the conductivity when equilibrated back at $25^o$C. Reaction (i) involved  0.3mL of the reactant  which was dissolved in 50mL of ethanol initially. The reaction was conducted at $25^o\mbox{C}$ and  monitored over time (minutes) by measuring conductivity ($\mu\text{S cm}^{-1}$) due to the release of $\text{H}^+$ and $\text{Cl}^-$ ions as shown below in  (\ref{eq:1}),
 \begin{equation} \label{eq:1}
\mbox{C$_4$H$_9$Cl}+\mbox{H$_2$O}\,\, \stackrel{k_a}{\longrightarrow}\,\,  \mbox{C$_4$H$_9$OH} + \mbox{H}^{+} + \mbox{Cl}^{-}.
\end{equation}

\item the methanolysis of ionized phenyl salicylate derived from the literature \cite[Table 7.1,p.381]{khan2} with presumably  accurate values of both the initial concentration and for all data sets of the kinetic run. 
Reaction (ii)  may be written
 \begin{equation} \label{eq:1b}
\mbox{PS}^{-}+\mbox{CH$_3$OH}\,\, \stackrel{k_b}{\longrightarrow}\,\,  \mbox{MS}^{-}  + \mbox{PhOH} 
\end{equation}
 where for  the rate law is pseudo first-order  expressed as  
\[\mbox{rate}=k_b\mbox{[PS]}^{-}=k_c[\mbox{CH$_3$OH}][\mbox{PS}^{-}].\]  
The methanol concentration is in excess and is effectively constant for the reaction runs \cite[p.407]{khan2}. The data for this reaction is given in detail in \cite[Table 7.1]{khan2}, conducted at $30^o\text{C}$ where several ionic species are present in the  reaction solution from  KOH, KCl, and $\text{H}_2\text{O}$ electrolytes.

\item the primarily SN1 substitution reaction  \cite[Table IX,p.2071]{ingold1} of tertiary butyl bromide ($\mbox{Bu}^t\mbox{Br}$) with dilute ethyl alcoholic sodium ethoxide in ethanol solvent where there concurrently occurs an approximately 20\%  contribution of an E1 elimination reaction.Reaction (iii) may be written 
 \begin{equation} \label{eq:1c}
\mbox{CH}_3\mbox{CBr}\,\, \stackrel{SN1}{\xrightarrow{\hspace*{1cm}}}\,\,  \mbox{Products} 
\end{equation}
where the solvent was EtOH with initial  sodium ethoxide concentration $[\mbox{NaOEt}]=0.02386$N at $25^o$C. The products consisted of approximately 81\% substituted tertiary butyl ethoxide and 19\% olefinic molecules  due to E1 elimination. Hence the rate constant here refers to a composite reaction (details in \cite[2064]{ingold3}  and \cite[p.2070]{ingold1}.

\item the second order E2  elimination reaction \cite[p.2059-2060 and Table VII,p.2064]{ingold3} with reactants isopropyl bromide ($\mbox{Pr}^i\mbox{Br}$) and sodium ethoxide (NaOEt). Reaction (iv) involving isopropyl bromide $\mbox{Pr$^i$Br}\equiv \mbox{(CH$_3$)CHBr(CH$_3$)} $ may be written 
\begin{equation} \label{eq:1d}
\mbox{(CH$_3$)CHBr(CH$_3$)} \,+ \, \mbox{OEt$^-$}\,\, \stackrel{E2}{\xrightarrow{\hspace*{1cm}}} \mbox{CH$_2$CHCH$_3$}
+ \mbox{Br$^-$  + HOEt}
\end{equation}
where the isopropyl bromide reacts with the \mbox{OEt$^-$} ion in EtOH solvent at \mbox{25$^o$C} to yield 80.3\% of the olefinic product  with some SN2 substitution  with the (OEt) functional group \cite[Table III,p.2061]{ingold3} according to the kineticists. Further details and data appear \cite[Table VII,p.2064]{ingold3}. It should be mentioned that the E2 reaction was inferred to be second order from prior experimental considerations since the [NaOEt] concentration reduction from the data exactly coincides with the reduction of \mbox{Pr$^i$Br} and was not independently determined. This perhaps somewhat experimentally questionable technique may well be the reason for the instantaneous rate constant  as computed here to be somewhat non-smooth, as will be discussed later  (see Fig.(\ref{fig:14})  for the  graph).
\end{enumerate}

``Units'' in the figures and text pertain to the appropriate reaction variable dimension, for instance either the absorbance for (i) or  the conductivity   ($\mu\text{S cm}^{-1}$) for (ii) below. Either because of evaporation or the temperatures not equilibrating after heating, the measured $\lambda_\infty$, it would be inferred that for (i) the measured value is larger than the actual one determined from the analysis.
Reaction (ii) is very rapid compared to (i) and the experimental data  plots show high nonlinearity. We denote by $\lambda $ the measurement parameter which is the conductivity $\mu\text{S cm}^{-1}$  or absorbance $A$  \cite[eqn 7.24-7.26]{khan2} for reactions (i) and (ii) respectively; $\lambda$ also refers to the concentration [X] of species X for reactions (iii) and (iv).    The more accurately determined $\lambda_\infty=A_\infty$ for (ii) \cite[Table 7.1,p.381]{khan2} was at approximately $0.897$.   Analysis of (ii) give values of $A_\infty =\lambda_\infty $ very close to the experimental ones that suggests that our determination  for reaction (i) $\lambda_\infty$ is correct. The experimental data and number of readings for the determination of rate constants is always related to the method used and the order of accuracy required in the study; for Khan \cite[Table 7.1,first $\mbox{A}$ column]{khan2}, (reaction (ii)) , 14 normal readings over 360 seconds (s) sufficed for Khan's purposes, whereas for the practical class (reaction (i)), 36 readings over 55 minutes (mins)  were taken.  The meager 14 readings of (ii) covered a major portion of the nonlinear region of the reaction, whereas for (i) the many readings were confined to the near-linear regime.
 Linear proportionality is assumed between $\lambda$  and the extent of reaction $x$, where  the first order  law (c being  the instantaneous concentration, $k$ the general rate constant and $a$ the initial concentration)   is $\frac{dc}{dt}=-kc=-k(a-x)$; with $\lambda_{\infty}=\alpha a,\lambda_{t}=\alpha x$ and $\lambda(0)=\lambda_{0}=\alpha x_0$, integration yields for assumed constant $k$
\begin{equation} \label{eq:2}
\ln\frac{(\lambda_\infty- \lambda_0)}{(\lambda_\infty-\lambda(t))}=kt
\end{equation}
Eqn.(\ref{eq:2}) determines $k$ if $\lambda_{0}$ and $\lambda_{\infty}$  are known.

   The analysis for  the latter  reactions (ii)-(iv) would provide a reference and indication of the predicted value in (i) for the initial concentration, apart from checking for overall consistency of the methodology in general situations especially when there is doubt concerning the value of the initial concentration.

The methods presented here   applies to any order provided the expressions can be expanded as  an $n$-order polynomial of the concentration variable against the time independent variable. To get smooth curves that are stable one had to modify and use a proper curve-interpolation technique that is stable which does not form sudden kinks or points of inflexion  and this follows next.

\subsection{Orthogonal polynomial stabilization}

It was discovered that the usual least squares polynomial method using Gaussian elimination \cite[Sec.6.2.4,p.318 ]{yak1}to derive the coefficients of the polynomial was highly unstable for $npoly>4$, which is a known condition \cite[p.318,Sec 6.2.4]{yak1}.  For higher orders, there is in addition the tendency to form kinks and loops  in an interpolated curve for values between two known intervals. Other methods described  in specialized  treatises \cite[Ch.5, Sec.5.7-5.13]{nrc} , even if robust and stable,such as the Chebyshev approximation required values of the proposed experimental curve at predetermined definite points in time, which is outside  the control of one using predetermined data  and so for this  work, the least square approximation was stabilized by  orthogonal polynomials \cite[Sec.6.3]{yak1} modified for determination of differentials. It is hoped that the method can also be extended  to integrals in future investigations.  The usual method defines the $n^{th}$ order polynomial $p_n(t)$ which is then expressed as a sum of square terms over the domain of measurement to yield $Q$ in (\ref{eq:5}). 
\begin{equation} \label{eq:5}
\begin{array}{rcl}
p_n(t) &=& \sum_{j=0}^{n}h_jt^j\\[5mm]
Q(f,p_n) &=&  \sum_{i=1}^N\left[f_i-p_n(t_i)\right]^2
	\end{array}
\end{equation}
The $Q$ function is minimized over the polynomial coefficient space. In the orthogonal method adopted here, we  express  our polynomial expression $p_m(t)$ linearly in  coefficients $a_j$  of $\varphi_j$  functions that are orthogonal with respect to an {\it inner} product definition. For  arbitrary functions $f,g$, the inner product $(f,g)$  is defined below, together with properties of the $\varphi_j$ orthogonal polynomials:  

\begin{equation} \label{eq:6}
\begin{array}{rcl}
(f,g) &=& \sum_{k=1}^{N} f(t_k).g(t_k)\\[2mm]
 (\varphi_i,\varphi_j)=0  &(i\neq j)&  \,\, \mbox{and} \,\, (\varphi_i,\varphi_i)\neq 0.
	\end{array}
\end{equation}

\begin{equation} \label{eq:7}
\begin{array}{rll}
\varphi_i(t)&=&(t-b_i)\varphi_{i-1}(t)-c_i\varphi_{i-2}(t)\,(i\geq1),\\[2mm]
 \varphi_0(t)&=&1,\,\, \mbox{and}\,\, \varphi_j=0  \,\,\mbox{for}\,\, j<1,\\[2mm]
 b_i &=& (t\varphi_{i-1},\varphi_{i-1})/( \varphi_{i-1}, \varphi_{i-1}) \,\, (i\geq1),b_i=0\, (i<1), \\[2mm]
 c_i &=&(t\varphi_{i-1},\varphi_{i-2})/( \varphi_{i-2}, \varphi_{i-2}) \,\, (i\geq2),\mbox{and}\, c_i=0\,\,(i<2). 
	\end{array}
\end{equation}
We define the $m^{th}$ order polynomial  and associated $a_j$ coefficients as follows:

\begin{equation} \label{eq:8}
\begin{array}{rll}
p_m(t)&=& \sum_{j=0}^m a_j\varphi_{j}(t)\\[0.5cm]
a_j &=&(f, \varphi_{j})/(\varphi_{j}, \varphi_{j}), (j=0,1,\ldots m) \\
 	\end{array}
\end{equation}

The recursive definitions for the first and second derivatives are given respectively as:
\begin{equation} \label{eq:9}
\begin{array}{rll}
\varphi^{\prime}_i(t)&=&\varphi^{\prime}_{i-1}(t) (t-b_i)+\varphi_{i-1}(t)-c_i\varphi^{\prime}_{i-2}(t)\,(i\geq1)\\[5mm]
\varphi^{\prime\prime}_i(t)&=&\varphi^{\prime\prime}_{i-1}(t) (t-b_i)+2\varphi_{i-1}^{\prime}(t)-c_i\varphi^{\prime\prime}_{i-2}(t)\,(i\geq2)
\end{array}
\end{equation}
Here the codes were developed in C/C++ which provides for recursive functions which we exploited for the evaluation of all the terms. The experimental data were fitted to an $m^{th}$ order expression  $\lambda_m(t)$ defined below 
\begin{equation} \label{eq:10}
\lambda_m(t) = \sum_{j=0}^{m}h_jt^j= p_m(t) = \sum^m_{j=0}a_j\,\varphi_j(t)\\
\end{equation}

The coefficients $h_i$ are all computed recursively, and the derivatives determined from (\ref{eq:10}) or from (\ref{eq:8}) and (\ref{eq:9}). Once $h_j$ or $a_j$ are determined, then the gradient  to the curve $\lambda_m(t)$ is computed as 
\begin{equation}
\lambda^\prime_m(t)=\sum_{j=0}^m jh_jt^{j-1}.
\end{equation}
The $l_Q(t)$ function of (\ref{eq:15}) is expanded similarly as for $\lambda_m(t)$ for order $m$.
 The orthogonal polynomial method is stable  and the mean square error decreases with higher polynomial order in general  monotonically (where $n$ is used to denote the integer order), but the differentials are not so stable, because of the contribution of higher order coefficients in the differential expression as will be shown. From the form of the of the equation that will be developed, the rate constant is determined as the  gradient of a straight-line graph in the appropriate segment of the graph. However, the curvature of the plot will increase with increasing $n$, giving a poorer value of $k$, whereas higher values of $n$ would better fit the $\lambda \, vs\, t$ curve. Hence inspection of the plots is  necessary to decide on the appropriate $n$ value, where we choose the lowest $n$ value for the most linear graph of the expression under consideration   that also provides a good $\lambda(t)$ fit over a suitable time range over which the $k$ rate constants apply. The orthogonal polynomial stabilization method provides good  $\lambda$ fits with increasing $n$, but not gradients, so that the onset of sudden changes to the gradient which  on physical grounds is unreasonable can be used as an indication as which is the best curve  to select. There is  in practice little  ambiguity in selecting the appropriate polynomials, as will be demonstrated.  Reactions (i)  and (ii)  both gauge  initial concentrations in terms  of the $A_\infty $ ( $\lambda_\infty$)  or final reading of a physical factor proportional to concentration  and the structure of the analysis  is the same and will therefore  be discussed  simultaneously, followed by reactions (iii) to (iv) , where concentrations are measured directly during the course of the reaction, which will be discussed together because the form of the boundary conditions and data are of the same class.  
\section{ANALYSIS OF REACTIONS (I) and (II)}
\begin{figure}[htbp]
\begin{center}
\includegraphics[width=7cm]{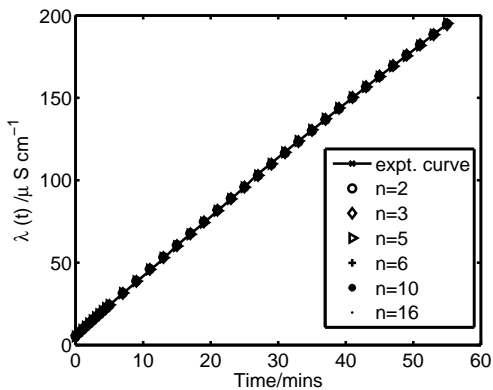} 
\end{center}
\caption{Plot of (i) using orthogonal polynomials for various orders $n$. The the least squares deviation goes down dramatically with increasing $n$, which was found not to be the case with the normal non-orthogonal polynomial method.}
\label{fig:4} 
\end{figure}

Figure (\ref{fig:4}) are plots for the different polynomial orders $n$ for reaction (i). It will be noticed that higher $n$ values in general leads to better fits visually; the normal least squares method leads to severe kinks and loop formation for $\sim n \geq 4$ which is not evident here. The reaction (ii) data covers a far greater domain with respect to  half-lifetimes with only about 14 points (which is a poor dataset with respect to our methods but which still gives quantitatively accurate values); because of the relatively more rapid curvature changes, we would expect very different gradient behavior as compared to (i) with its stronger linearity.

\begin{figure}[htbp]
\begin{center}
\includegraphics[width=7cm]{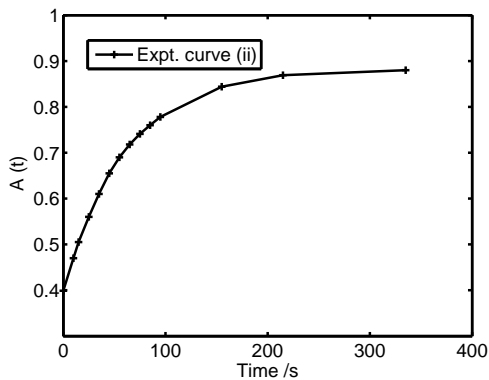} 
\end{center}
\caption{Experimental points omitting point at $A_\infty$ for reaction (ii) at time $=2135s$. The curve is rather non-linear.}
\label{fig:c2exp} 
\end{figure}

The corresponding plots for reaction (ii) are  in Fig. (\ref{fig:c2exp}).

\begin{figure}[htbp]
\begin{center}
\includegraphics[width=7cm]{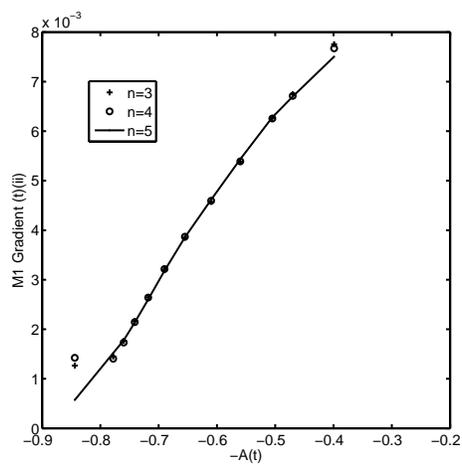} 
\end{center}
\caption{Method 1 applied to reaction (ii). Only at the peripheral value does the 
fit fail for lower values of n due to the extreme curvature.}
\label{fig:m1react2} 
\end{figure}

In view of the nonlinearity, we chose a limited regime to curve fit for polynomial order $n=3,4,5$ in Fig.(\ref{fig:c2exp}) and the gradient  was computed for the $n=5$ polynomial to determine the rate constants as it was the only order that gave a smooth curve for the  first $12$ consecutive points in the range; the other orders  also gave consistent and almost equal gradients except at the extreme end points of the range plotted as  depicted for example  in Figs. ( \ref{fig:m1react2}, \ref{fig:m2react2}, \ref{fig:5c}). 
\begin{figure}[htbp]
\begin{center}
\includegraphics[width=7cm]{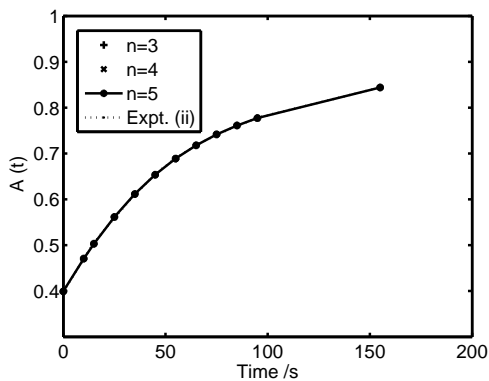} 
\end{center}
\caption{Experimental points curve fitted with polynomials of order $n=3,4,5$. The fit for this range is excellent, despite the nonlinear nature of the curve}
\label{fig:c2n35fit}
\end{figure}

\begin{figure}[htbp]
\begin{center}
\includegraphics[ width=7cm]{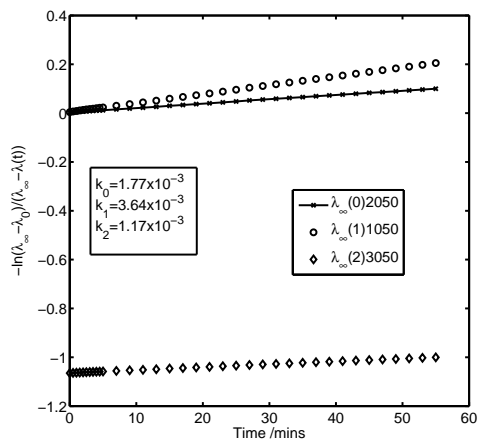} 
\end{center}
\caption{Integrated equation(\ref{eq:2}) plot with $\lambda_{\infty}$ from experiment (0) and from two different arbitrary values (1,2) for $\lambda_{\infty}$, which yields two different values for the rate constant due to gradient change.}
\label{fig:1} 
\end{figure}
Unlike reaction (ii), the $\lambda_{\infty }$ for reaction (i) was ambiguous.  The plot of (\ref{eq:2}) was made for  the same experimental values with different $\lambda_\infty$'s, both higher and lower than the supposed experimental value for this reaction. The plots  in Fig.(\ref{fig:1})shows increasing $k_a$ for decreasing $\lambda_{\infty}$; the choice $\lambda_\infty=1050$  leads to a value  of $k_a$  close to the NLA values for the different methods discussed which  does not require $\lambda_\infty$, but is also able to determine this value by extrapolation. The rate constant from  NLA  is higher than that determined from experiment, implying a lower $\lambda_\infty$  value  which is consonant with evaporation of solvent and/or the non-equilibration of temperature prior to measurement  to determine $\lambda_{\infty}$. Hence elementary NLA allows one to deduce the accuracy of the actual experimental methodology in this example. Except for one section, we shall apply  NLA based on constant $k$ assumption. We also quote some values of Khan's results \cite[Table 7.1]{khan2} in Table (\ref{tab7p1}), where some comment is required. The $A$ absorbance is monotonically increasing and at higher time ($t$) values (see \cite[Table 7.1]{khan2}) the experimental $A$ value exceeds the $A_\infty$  that is determined by the process of minimizing $\sum d_i^2$. Hence the minimization of $\sum d_i^2$ with respect to $A_\infty$ is taken as a protocol for determining the best $k$ value even if it contradicts experimental observation. Further, this protocol is highly sensitive to $A$; a change of $10^{-3}$ leads to an approximately  tenfold change in $k$. On the other hand, if $A_\infty$ determined from experiment as $0.897$ is accepted, then then computed rate constant for this value is $k=2.69\times 10^{-3}\mbox{s}^{-1}$ implying that the uncertainty in $k$ is of the order of $\pm 14 \times 10^{-3}$. Hence we can conclude  that the Khan method is a protocol that accepts as correct  the $k$ value that is determined by the minimization of $A_\infty$  for a certain $A_\infty$ range ($\approx 0.8980-.8805$), which again refers to an unspecified protocol as to the choice of the range.

\begin{table}
\begin{center}
\begin{tabular}{|l|c|c|c|c|c|c|} \hline
$10^3\sum d_i^2$&513.5&109.4&8.563&63.26&212.7&227.4\\
\hline
$A_\infty$   & .8805 & .881 & .882 & .883 & .885 & .887 \\
\hline
$10^3k/s^{-1}$  & $19.7\pm.6$ & $18.1\pm.3$ & $16.5\pm.1$ & $15.5\pm.2$ & $14.2\pm.4$ & $13.3\pm.5$ \\
\hline

\end{tabular}
\caption{Some results from reaction (ii) \cite[p.381,Table 7.1]{khan2}.The first row refers to the square difference summed, where the lowest value would in principle refer to the most accurate value (third entry from left). The second row refers to the $A_\infty$ absorbance  and the last to the corresponding rate constant with the most accurate believed at the stated units to be  at $16.5\pm.1$.}
\label{tab7p1}
\end{center}
\end{table}


\subparagraph{The general 1st and 2nd order equations}
We state  the standard integrated forms below  as a reference that requires specification of initial concentrations in order to contrast   them to the methods developed  here.

The first order rate constant $k_1$ is determined from 
\begin{equation} \label{eq:t1}
k_1=
		\frac{1}{t} \log_e\{b(b-x)\}
\end{equation}

 and the  second  order rate constant $k_2$ is determined from 
\begin{equation}\label{eq:t2}
	k_2=\left[
		\frac{1}{t(a-b)}
	\right]\log_e\frac{b(a-x)}{(b-x)}
\end{equation}
where  these  expressions are given in \cite[p.2063]{ingold3} and utilized to compute rate constants where $a$ and $b$ are the initial concentration terms  at $t=0$ and $x$ is the extent of reaction.

\subsubsection{Method 1} 
\begin{figure}[htbp]
\begin{center}
\includegraphics[width=7cm]{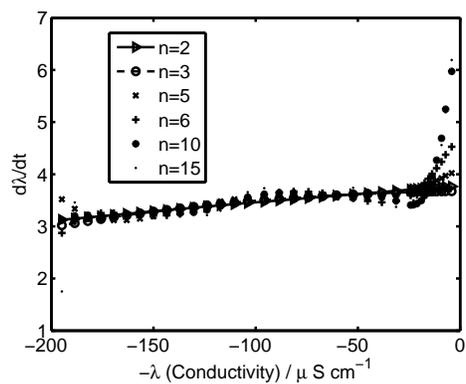} 
\end{center}
\caption{Method 1 graph showing  linearity  lower order polynomial fits for reaction (i).}
\label{fig:2} 
\end{figure}

This method is a variant of the direct method of eq.(\ref{eq:15}).
For constant $k$, the rate equation $\frac{dc}{dt}=-kc=-k(a-x)$ reduces to 
\begin{equation} \label{eq:2b}
\frac{\lambda(t)}{dt}= -k\lambda(t)+\lambda_{\infty}.k
\end{equation}
Hence a  plot of $\frac{\lambda(t)}{dt}$ vs $\lambda(t)$ would be linear. We find this to be the case for polynomial order $npoly\leq 3$ as in Fig.(\ref{fig:2}) below for all data values; higher polynomial orders can be used in selected data points  of the  curve below, especially in the central region. Thus criteria must be set up to determine the appropriate regime of data points for a particular polynomial order in  NLA. For n=2, $k_a=3.34\pm .03\times 10^{-3}\mbox{min}^{-1}$ and $\lambda_\infty =1134\pm10 \,\mbox{units}$.
The plots for reaction (ii) is  a little more involved; it is a much more rapid reaction and the  number of data-points are relatively sparse  for NLA and the points cover the entire range of the reaction sequence and is highly non-linear; it was found that the gradients were smooth for the first 10 or so points and reasonably linear, but that at the boundary of these selected points, there are deflections in the curve; on the other hand , the different order polynomial curves ($n\leq 5$) are all coincident over a significant range of these values; we chose the $n=5$ polynomial curve to determine the curve over the entire range and the linear least squares fit yields the following data $k_b=1.64\pm.04\times10^{-2}\mbox{s}^{-1}$
and $A_\infty =0.8787\pm .0008\,\mbox{units}$.

\subsubsection{Method 2}
This method is yet another  variant of the direct method of eq.(\ref{eq:15}).
Let $\alpha'=\lambda_\infty-\lambda_0$, then  $\ln \alpha' -\ln(\lambda_\infty-\lambda)=kt$, then noting this and differentiating  yields 
\begin{equation} \label{eq:3}
\underbrace{\ln\left(\frac{d\lambda}{dt}\right)}_{Y}=\underbrace{-kt}_{Mt}+\underbrace{\ln[k(\lambda_\infty- \lambda_0)]}_{C}
\end{equation}
\begin{figure}[htbp]
\begin{center}
\includegraphics[width=7cm]{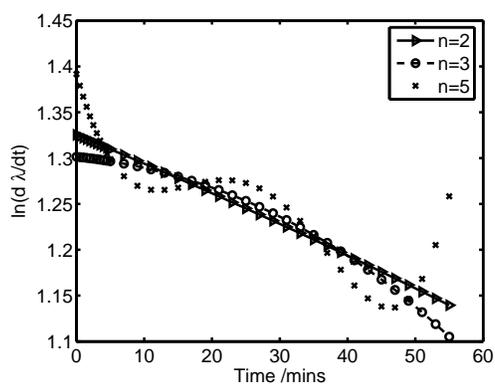} 
\end{center}
\caption{Method 2 reaction (i) where smooth curves are obtained for  at least $npoly<4$ .}
\label{fig:3} 
\end{figure}

\begin{figure}[htbp]
\begin{center}
\includegraphics[width=7cm]{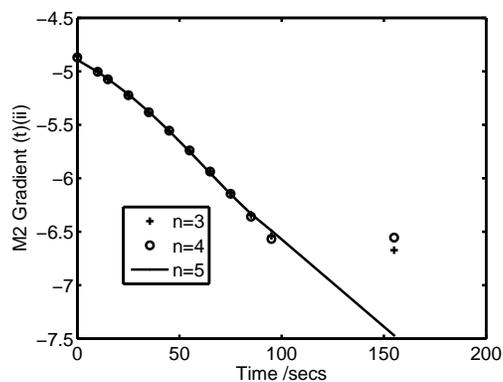} 
\end{center}
\caption{Method 2 reaction (ii) where smooth curves are obtained for  $npoly<5$. The $n=5$ polynomial is used to calculate the best linear line over this range as the other polynomials do not fit for the last value in the series due to the extreme curvature. }
\label{fig:m2react2} 
\end{figure}
A typical plot that can extract $k$ as a linear plot of $\ln(d\lambda/dt)$ vs $t$ is given in Fig.(\ref{fig:3}) for Method 2, reaction (i) and in Fig.(\ref{fig:m2react2}) for Method 2, reaction (ii).  Linearity  is observed  for $npoly=2$ and smooth curves without oscillations for at least $npoly\leq3$ for reaction (i) and an analysis for reaction (ii) uses $npoly=5$. The linear least square  line yields  for Method 2 the following:
 \[ \begin{array}{rc}
 k_a=&3.35\pm .03\times 10^{-3}\mbox{min}^{-1} \,\text{and}\, \lambda_\infty =1130\pm10 \,\mbox{units}\\

k_b=&1.72\pm.02\times10^{-2}\mbox{s}^{-1}
\,\text{and}\, A_\infty =0.86(53)\pm .02\,\mbox{units}.
 \end{array}
 \]

We note that because of the manifest nonlinearly of the gradients, one cannot determine the $A_\infty $ values to $4$-decimal place accuracy as quoted by Khan based on his model and assumptions \cite[Table 7.1]{khan2}.

\subsubsection{Other associated non-direct methods}
There are other methods, one of which is a variant of Method 1  and another that utilizes a least-squares optimization of the form of the equation for first and second derivatives. 
\subsubsection{Method 1 variation}\label{ss1}
A variant  method similar to the Guggenheim method \cite{gug1} of elimination is given below but where gradients to the conductivity curve is required, and where the average over all pairs is required; the equation follows from (\ref{eq:3}).
\begin{equation} \label{eq:4}
\left\langle k\right\rangle=\frac{-2}{N(N-1)}\sum_{i}^{N}\sum_{j>i}^N\ln\left(\lambda'(t_i)/\lambda'(t_j)\right)/
(t_i-t_j)
\end{equation}
Since we are averaging over instantaneous $k$ values, there would be a noticeable standard deviation in the results if the hypothesis of change of rate constant with species concentration is correct. Differentiating (\ref{eq:3}) for constant $k$ leads to (\ref{eq:11}) expressed in two ways
\begin{equation} \label{eq:11}
\frac{d^2 \lambda}{dt^2}=-k\left(\frac{d\lambda}{dt}\right)\,(a) \,\mbox{or}\,
k=-\frac{d^2 \lambda}{dt^2}/\left(\frac{d\lambda}{dt}\right)\,(b)
\end{equation}

If $\lambda(t)=\sum_{i=0}^{n+1 }a(i)t^{i-1}$, then as $t\rightarrow 0$, the rate constant is given by $k=\frac{-2a(2)}{a(1)}$ from (\ref{eq:11}b). For the above, $n, id, \,\text{and}\,iu$ denotes as usual the polynomial degree,the lower coordinate index and the upper  index of consecutive coordinate points respectively, where the average is over  the consecutive points, whereas the $k$ rate constant with subscript "all"  below refers to the equation (\ref{eq:4}) . 

The results from this calculation are as follows:
 \[ \begin{array}{c}
 k_{a,all},k_{a,id,iu}=3.32, 3.23\pm.07\times 10^{-3}, \text{min}^{-1}, \,\mbox{$n=2,id=10,iu=20$}\\
 k_{a,t\rightarrow 0}= 3.082\times 10^{-3}   \text{min}^{-1}.\\
 k_{b,all},k_{b,id,iu}=1.7150, 1.676\pm.3\times 10^{-2}, \text{s}^{-1},
 \,\mbox{$n=5,id=1,iu=10$}\\
 k_{b,t\rightarrow 0}= 1.023\times 10^{-2}   \text{s}^{-1}.
 \end{array}
 \]

The asymptotic limit gives a lower value for $k_b$  than for the other methods for reactions (i) and (ii). One possible explanation is that the rate constant changes as a function of time, but we note that (\ref{eq:11}) was derived assuming constant $k$.

\subsubsection{Optimization of first and second derivative expressions}\label{ss2}
 Eq.(\ref{eq:11}(b)) suggests another way of computing $k$ for ``well-behaved'' values of the differentials, meaning regions where $k$ would appear to be a reasonable constant. The (a) form suggests an exponential solution. Define $\frac{d\lambda}{dt}\equiv dl$ and $\frac{d^2\lambda}{dt^2}\equiv d2l$. Then $dl(t)=A\exp(-kt)$ and $dl(0)=A=h_2$ from (\ref{eq:10}).Furthermore,  as $t\rightarrow 0$, $k=\left(-2h_2/h_1\right)$ and a global definition of the rate constant becomes possible based on the total system $\lambda(t)$ curve.
 
With a slight change of notation, we now define $dl$ and $d2l$ as referring to the continuous functions $dl(t)=A\exp(-kt)$ and $d2l(t)=-kA\exp(-kt)$ and we consider $(d\lambda/dt)$ and $d^2\lambda/dt^2 $ to belong to the values (\ref{eq:10}) derived from ls fitting where $(d\lambda/dt)=\lambda_m^{\prime}$, \,\, $(d^2\lambda/dt^2)=\lambda_m^{\prime\prime}$ which are the  experimental values for a curve fit of order $m$. From the experimentally derived gradients and differentials, we can define two non-negative functions $R_\alpha(k)$ and $R_\beta(k)$ as below:
 \begin{equation} \label{eq:12}
 \begin{array}{rll}
 R_\alpha(k)&=&\sum^N_{i=1}\left(\frac{d^2\lambda(t_i)}{dt^2}+kdl(t_i)\right)^2\\[.3cm]
 R_\beta(k)&=&\sum^N_{i=1}\left(\frac{d\lambda(t_i)}{dt}-dl(t_i)\right)^2\\
 &\mbox{where}&  \\
 f_\alpha(k)=R^{\prime}_\alpha(k) &\mbox{and}& f_\beta(k)=R^{\prime}_\beta(k)
 \end{array}
\end{equation}
 and a stationary point (minimum) exists at $f_\alpha(k)=f_\beta(k)=0.$ We solve the equations $f_\alpha$ , $f_\beta$  for their roots in k using the Newton-Raphson method to compute the roots  as the rate constants $k_\alpha$ and $k_\beta$ for functions $f_\alpha(k)$ and $f_\beta(k)$ respectively. The error threshold in the Newton-Raphson method was set at $\epsilon=1.0\times10^{-7}$ We provide a series of data of the form $\left[n,A,k_\alpha,k_\beta ,\lambda_{\alpha,\infty},\lambda_{\beta,\infty}\right]$ where $n$ refers to the polynomial degree, $A$ the initial value constant as above, $k_\alpha$  and $k_\beta$ are the rate constants for  the functions $f_\alpha$ and $f_\beta$  (solved when the functions are zero respectively ) and 
 likewise  for $\lambda_{\alpha,\infty}$ and $\lambda_{\beta,\infty}$. The $e$ symbol refers to base $10$ (decimal) exponents. The $\lambda_{\infty}$ values are averaged over  all the 36 data points for reaction (i)  and for the 12 datapoints of reaction (ii) from  the equation 
 \begin{equation} \label{eq:12a}
 \lambda_{\infty}=\frac{d\lambda(t)}{dt}\frac{1}{k}+ \lambda(t)
\end{equation}
for scheme $\alpha$ and $\beta$ for both reaction (i) and (ii).
The results are as follows.\\
\textbf{Reaction (i)}\\
$\left[2, 3.7632\times10^0, 3.2876\times10^{-3}, 3.2967\times10^{-3}, 1.1506\times10^{3}, 1.1477\times10^{3} \right]$,\\
$\left[3, 3.6384\times10^0, 2.7537\times10^{-3}, 2.7849\times10^{-3}, 1.34756\times10^{3},1.3334\times10^{3}\right]$,\\
$\left[4, 3.6384\times10^0, 2.0973\times10^{-3}, 2.4716\times10^{-3}, 1.7408\times10^{3}, 1.4900\times10^{3}\right]$,\\
$\left[5, 4.0213\times10^0, 9.7622\times10^{-3}, 4.9932\times10^{-3}, 4.4709\times10^{2}, 7.9328\times10^{2},\right]$,\\
$\left[6, 4.5260\times10^0, 4.1270\times10^{-2}, 8.9257\times10^{-3}, 1.7101\times10^{2}, 4.8403\times10^{2}\right]$.\\
We noticed as in the previous cases that the most linear values occur for $1<n<4$. In this approach, we can use the $f_\alpha$ and $f_\beta$ function similarity of solution  for $k_\alpha$ and $k_\beta$ to determine the appropriate regime for a reasonable  solution. Here, we notice a sudden departure of similar value between $k_\alpha$ and $k_\beta$  (about 0.4 difference ) at $n=4$ and so we conclude that the probable average  ``rate constant'' is about the range given by the values spanning $n=2$ and $n=3$. Interestingly, the $\lambda_{\infty}$  values are approximately similar to the ones for method 1 and 2 for polynomial order  2  and 3 for reaction (i). More study with reliable data needs to be done in order to discern and select  appropriate criteria that can be applied to these non-linear methods. Because of the large number of datapoints in the linear range, $k_a$ and $k_b$ values are very compatible for $n=2,3$ where the $k_a$ determination involves double derivatives, which cannot be determined with accuracy unless a sufficient number of points is used.  
\\
\textbf{Reaction (ii)}\\
The results for this system are\\
$\left[5, 7.5045\times10^{-3}, 1.2855\times10^{-2}, 1.5497\times10^{-2}, .94352, .89247\right]$\\
for  the first 12 datapoints of the published data to time coordinate $155$ secs. For polynomial order 3,4 and the first 11 datapoints, where there
are no singularities in the curve we have\\
$\left[3, 7.7275\times10^{-3}, 1.4469\times10^{-2}, 1.6147\times10^{-2}, .91320, .88335\right]$\\
$\left[4, 7.4989\times10^{-3}, 1.3146\times10^{-2}, 1.5359\times10^{-2}, .94208, .89652\right]$.\\
Here,  $k_a$ and $k_b$ differ by $\sim.2\times10^2 \text{s}^{-1}$; one possible reason for this discrepancy  is the insufficient number  of datapoints to to accurately determine $\frac{d^2\lambda}{dt^2}$.  Even if the number of points are large, experimental fluctuations would induce changes in the second derivative which would  be one reason for discrepancies.  Hence experimentalists who wish to employ NLA must provide more experimental points, especially at the linear region of the $\lambda(t)\, vs\, t$ curve.   
\subsection{Inverse Calculation}
Rarely are experimental curves compared with the ones that must obtain from the kinetic calculations. Since the kinetic data is the ultimate basis for deciding on values of the kinetic parameters, replotting the curves with the calculated parameters  to obtain the most fitting curve to experiment would serve as one method to determine the best method amongst several. For reaction (i) we have the following data:

 \begin{table}[htbp]
\begin{center}
\begin{tabular}{l|c|c|c|c}
Result & Procedure & Poly. order & $\lambda_\infty$ & $k_a$ \\
\hline
1 & From expt & - & 2050. & $1.7752\times10^{-3}$ \\
\hline
2 & Method 1 & 2 & 1134.3 & $3.3397\times10^{-3}$ \\
\hline
3 & Method 2 & 2 & 1130.23 & $3.347\times10^{-3}$ \\
\hline
4 & sec(\ref{ss2}) $R_\alpha$ & 2 & 1150.63 & $3.288\times10^{-3}$ \\
\hline
\end{tabular}
\caption[smallcaption]{Data for the plot of Fig.(\ref{fig:6}) for reaction(i)}
\label{tab:2}
\end{center}
\end{table}

For reaction(ii), we have the following:
 \begin{table}[htbp]
\begin{center}
\begin{tabular}{l|c|c|c|c}
Result & Procedure & Poly. order & $\lambda_\infty$ & $k_a$ \\
\hline
1 & From expt & - & .8820 & $1.65\times10^{-2}$ \\
\hline
2 & Method 1 & 5 & .8787 & $1.64\times10^{-2}$ \\
\hline
3 & Method 2 & 5 & .8653 & $1.72\times10^{-2}$ \\
\hline
4 & sec(\ref{ss2}) $R_\beta$ & 5 & .89247 & $1.5497\times10^{-2}$ \\
\hline
\end{tabular}
\caption[smallcaption]{Data for the plot of Fig.(\ref{fig:7})  for reaction (ii). }
\label{tab:3}
\end{center}
\end{table}
\begin{figure}[htbp]
\begin{center}
\includegraphics[width=9cm]{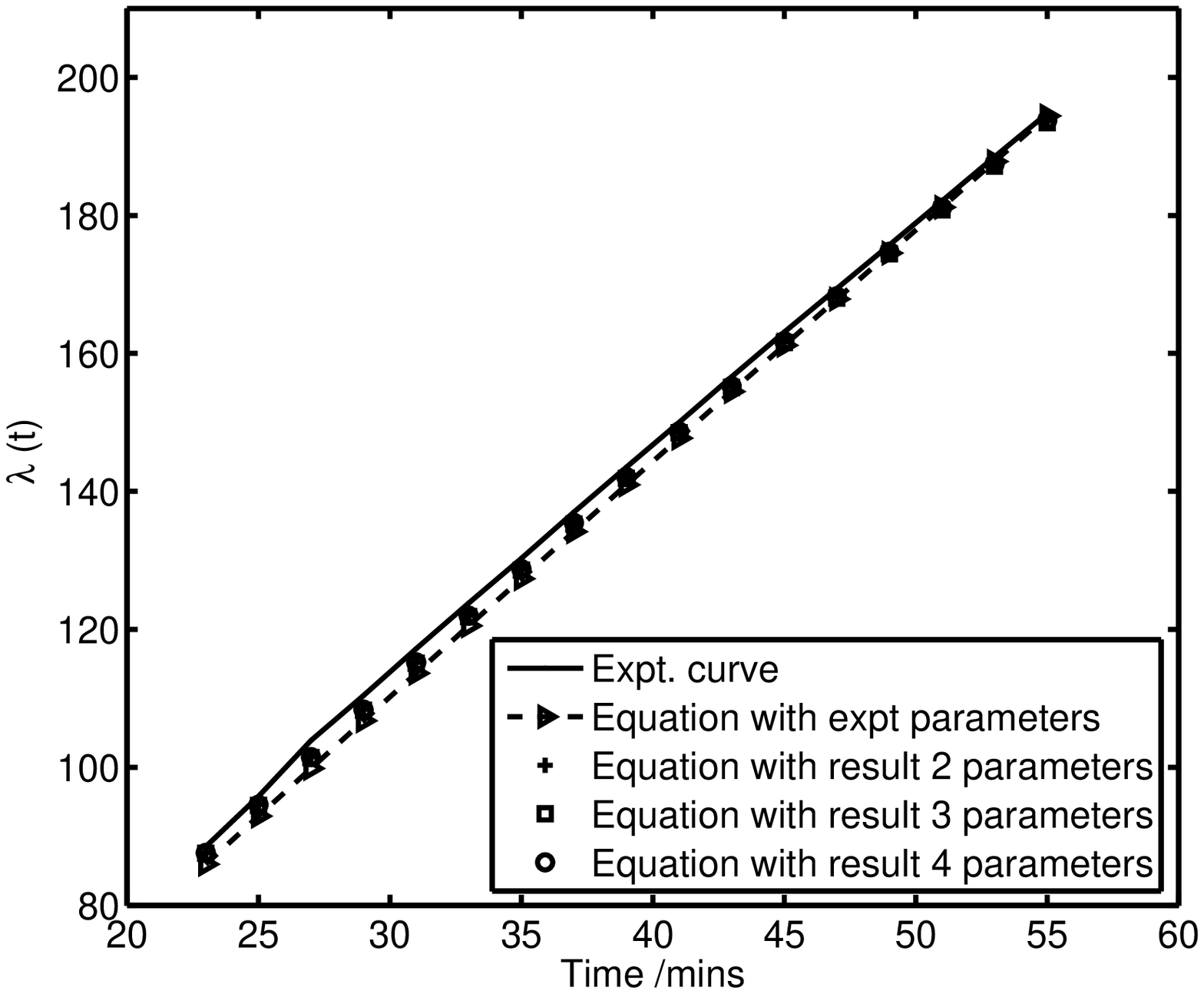} 
\end{center}
\caption{The plots according to   the $k_b$ and $A_infty$ values of Table(\ref{tab:2})}
\label{fig:6} 
\end{figure}

\begin{figure}[htbp]
\begin{center}
\includegraphics[width=9cm]{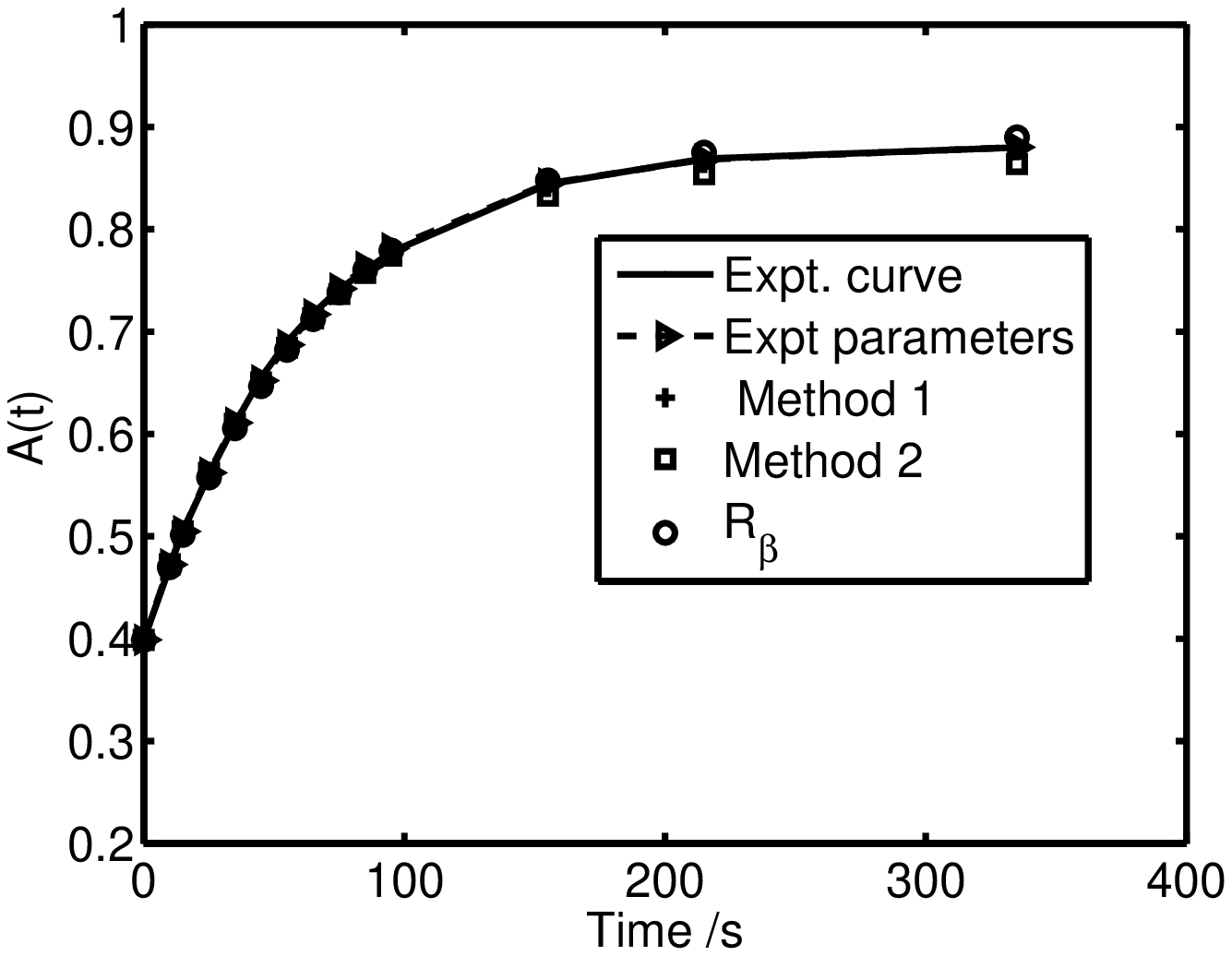} 
\end{center}
\caption{The plots according to   the $k_b$ and $A_infty$ values of Table(\ref{tab:3})}
\label{fig:7} 
\end{figure}

Fig.(\ref{fig:6}) indicate  that the parameters derived from experiment is the worst fit compared to the methods  developed here, verifying that our computations, including the $\lambda_\infty$ values are a better fit than the one derived from experiment due to flaws in the methodology of driving the reaction to completion by heating, leading to evaporation and therefore inaccurate determination. Based on the comparisons between the reactions (i) and (ii), we predict that reaction (i) if carried out under stringently controlled conditions, especially  in determining $\lambda_\infty $ would have a rate constant approximately  $\sim 3.2\times10^{-3}\text{mins}^{-1}$ rather than the experimentally deduced $\sim 1.77\times10^{-3}\text{mins}^{-1}$ with $\lambda_\infty\sim 1130$ units rather than $2050$ units . For reaction (ii), we note a good fit for all the curves, that of the experiment, Khan's results and ours.

\subsection{Evidence of varying kinetic coefficient $k$ for reactions (i) and (ii)} 
\begin{figure}[htbp]
 \begin{center}
	\includegraphics[width=7cm]{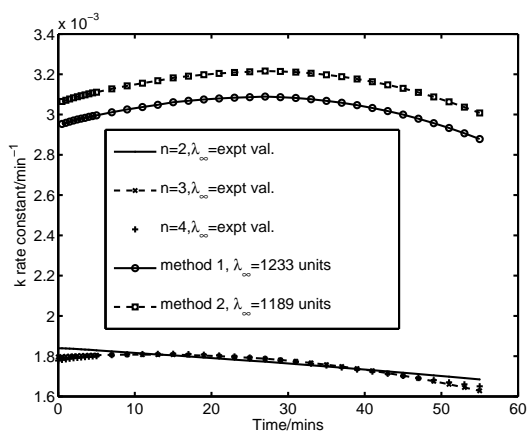}
	\end{center}
		
	\caption{Variation of $k$ with time or concentration changes based on the experimental value $\lambda_{\infty}=2050 \,\text{units}$ and the computations based on different polynomial degrees $n=2,3,4$ and the computed $\lambda_{\infty}$ values for Method 1 and Method 2 for fixed polynomial degree $n=3$.}
	\label{fig:5}
\end{figure}

Finally, what of direct methods that  do not assume the constancy of $k$ which was the case in the above subsections? Under the linearity assumption $x=\alpha\lambda(t)$, the rate law  has the form $dc/dt=-k(t)(a-x)$ where $k(t)$ is the instantaneous rate constant and this form implies 
 \begin{equation} \label{eq:13}
 k(t)=\frac{d\lambda/dt}{(\lambda_{\infty}-\lambda(t) )}
\end{equation}
 If $\lambda_\infty$ is known from accurate experiments or from our computed estimates, then $k(t)$ is determined; the variation of $k(t)$ could  provide crucial information concerning reaction kinetic mechanism and energetics,  from at least one theory  recently developed for elementary reactions \cite{cgj1}  \textit{at equilibrium}; and for such similar theories \cite{cgj7} and experimental developments for very large changes in concentration, it may be anticipated  that nonlinear methods would be used to accurately determine $k(t)$ that would yield the so-called ``reactivity coefficients'' \cite{cgj1} that account for variations in $k$ that would provide fundamental information concerning activation and free energy changes. However, since these coefficients pertain to the steady-state scenario where a precise relationship exists between the ratio of these coefficients and that of the activity, one might not expect to detect these coefficients for relatively minute concentration changes that occur in most routine chemical reaction determinations. On the other hand, the very preliminary results here seem to  indicate transient variations belonging possibly to another class of  phenomena ; it could well be due to  periodicity in the reactions  where some of the "beats" detected  -assuming no experimental error in the data-  could be related to the time interval between measurements, that is,  because the number of datapoints is restricted, only certain beats are observed in the periodicity. The assumption  that the spectroscopic detector is not noisy relative to the magnitude of the experimental data and that it does not have  significant periodic drift relative to a reference absorbance leads to the conclusion that some form of chemically induced periodicity might be present. It would  be very interesting to increase the number of datapoints where  the time interval between measurements is reduced and to analyze the different types of apparent frequencies that might be observed with different time intervals of measurement. From these observations, perhaps theories could be adduced on  the nature of these presumed fluctuations.         \\
\textbf{Reaction (i) results}\\
 Figure(\ref{fig:5}) refers to the computations under the assumption of first order linearity of concentration and the conductivity. Whilst very preliminary, non-constancy of the rate constants are evident, and one can therefore expect that another area of fruitful experimental and theoretical development can be expected from these results that incorporates at least these two effects.\\
 \textbf{Reaction (ii) results}\\

\begin{figure}[htbp]
 \begin{center}
	\includegraphics[width=7cm]{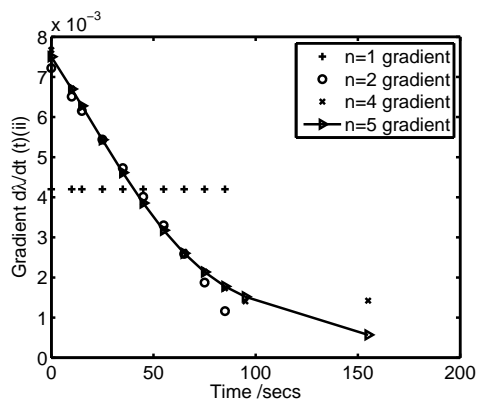}
	\end{center}
		
	\caption{Variation of gradient $d\lambda=dt=dA/dt$ with time for different polynomial orders $n=1,2,4 \,\text{and}\,5$}
	\label{fig:5c}
\end{figure}

To verify that the curious  results are not due to  minute fluctuations of the gradient, we plot the gradient $\frac{dl}{dt}$ of the curve fits for polynomial orders $1,2,4 \,\text{and}\,5$. Even for low orders, the fit is very good with no  oscillations observed between lower and higher order polynomials for approximately the first 10 values of the kinetic data in Fig.(\ref{fig:5c}). It was found that the lower order polynomials gave essentially the same results for the restricted domain  where the gradients coincided with those of higher order.  

\begin{figure}[htbp]
	\begin{center}
	\includegraphics[width=7cm]{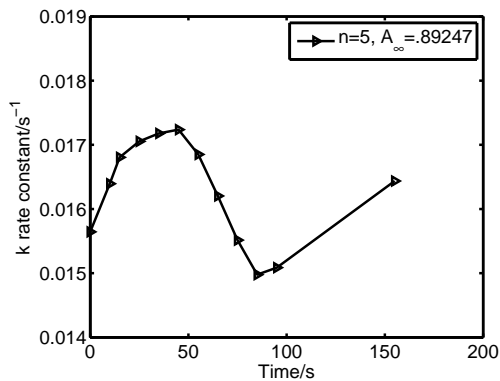} 
	\end{center}
	\caption{An apparently periodically varying rate constant that settles to a higher value  at larger time increments. $A_\infty $  was the value taken from Method 1 above for polynomial order $5$. }
	\label{fig:5b} 
	\end{figure}

The gradient drops to $0$ at  the long time $t\rightarrow \infty $ limit; on the other hand, the factor $\frac{1}{(\lambda_{\infty}-\lambda(t))}$ rises to infinity; so we might expect from these two competing factors various sinusoidal-like properties, or even maxima.    The surprising result is shown in Fig. (\ref{fig:5b}). It could be that  the form (\ref{eq:13}) is not valid because no instantaneous value of the rate constant can be defined. Also, it is not possible at this stage to definitively rule out the detector causing a sinusoidal variation due to  periodic drift and instability. On the other hand, if we can rule out artifacts due to systematic instrument error,  then one must admit the  possibility that a long-time cooperative effect involving coupling of the reactant molecules through the solvent matrix over the entire reaction chamber may take place . This seems like a big idea that could  be investigated further. However, if the results are due solely to instrument error, then this method allows us to monitor the  error fluctuations by taking a suitable weighted average.  We note that reaction (ii) is relatively complex, involving many ionic species and some intermediate steps or reactions \cite[p.414-416]{khan2}. This fact, coupled with the cell setup where steady state temperature gradients might well exist, would lead to coupled processes described by irreversible thermodynamics, which could possibly explain such rate constant changes relative to the first order parametrization used here.

\textbf{Comment:}
Barring artifacts, Figs.(\ref{fig:5}-\ref{fig:5b}) is consonant with two separate effects: ($\alpha$) a long-time limit due to changes in concentration that alters the force fields and consequently the mean rate constant value (according to the theories in \cite{cgj1, cgj7}) of the reaction as equilibrium is reached, and ($\beta$) possible transient effects due to collective modes of the coupling between the reacting molecules and the bulk solution as observed in the region between the start of the reaction and the long-time interval. In both reactions (i) and (ii), there appears a slight change in the rate constant value at time $t=0$ and the values at the end of the  experimental measured interval which may be due  to the altered force fields that would change slightly the rate constant according to ($\alpha$). On the other hand, there is a relatively slow  and minute  sinusoidal-like change in the rate constant that may be due to some cooperative effect, if no artifacts are implied; the interpolation with different polynomials leading to the same gradient seems to suggest that some type of collective behavior might be operating during the course of the reaction; if this is so then ($\beta$) would be a new type of phenomena that has not hitherto been incorporated into chemical kinetics research.

\section{RESULTS FOR REACTION (III) and (IV)} 
Two different methods are utilized to determine the reaction rate constants. The direct method utilizes determining the gradient $k$ of the $d[A_1]/dt \, vs \, [Q]$ curve of eq.(\ref{eq:15}) by fitting the best straight line. Initial concentrations are not required, and the error in the gradient may be estimated from the mean least squares error  of the end-points; define the mean square  per point $\Delta^2$ as  $\Delta^2=\sum^n_{i=1}(\bar{y}-y_i)^2/2$ where $\bar{y}$ is the linear optimized curve and $y_i$ a datapoint within the range of measurement. Then for the range of datapoints  $|X|=|Q|$ we estimate the error in the rate constant as $\Delta k= \sqrt{\Delta^2}/|X|$. For what follows below  the first order rate constant for reaction (iii) is denoted $k_{1d}$ derived from direct computation of the gradient of (\ref{eq:15}) whereas $k_{1ls}$ denotes the rate constant as calculated by our least squares method (\ref{eq:16}). Similarly, for reaction (iv), $k_{2d}$ denotes  the rate constant derived directly from the gradient of the curve following eq.(\ref{eq:15}) by fitting the best straight line and $k_{2ls}$ is the  second order rate constant from the \emph{least squares minimization} of (\ref{eq:16}). A detailed description follows below.

\subsection{Reaction (iii) first order details}
\begin{figure}[htbp]
\begin{center}
\includegraphics[width=7cm]{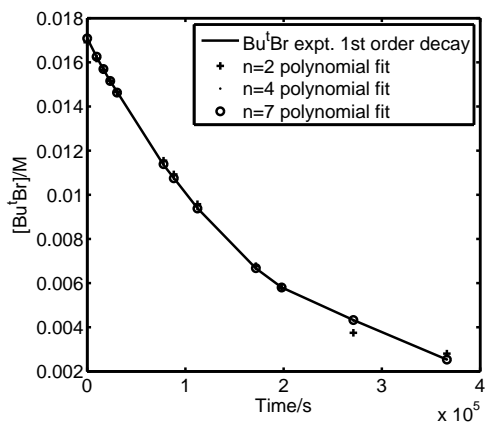} 
\end{center}
\caption{The polynomial degree $n$ for the orthogonal polynomial fit for  the first order $\mbox{Bu}^t\mbox{Cl}$  reaction (iii)  where there is near coincidence for $n=4,5,6,7$. }
\label{fig:8} 
\end{figure}

\begin{figure}[htbp]
\begin{center}
\includegraphics[width=7cm]{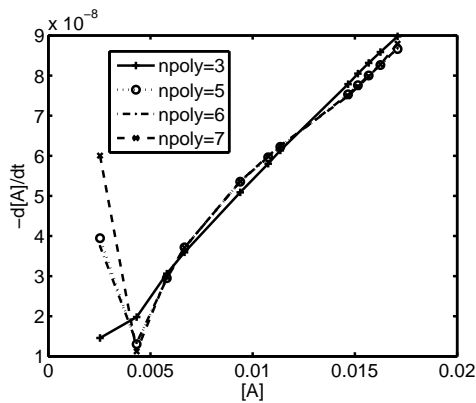} 
\end{center}
\caption{A plot of the rate of decomposition of $\mbox{Bu}^t\mbox{Br}(\equiv \mbox{A}) vs \mbox{A} $ where [A]  refers to the molar concentration of $\mbox{Bu}^t\mbox{Br}$ according to the kinetic data published in \cite[Table IX.]{ingold1} for reaction (iii). }
\label{fig:9} 
\end{figure}

Fig.(\ref{fig:8}) is a plot of the various orthogonal polynomial order fits. The $n=2$ order is rather poor but higher orders  all coincide with the experimental points. And for $n>2$, we find that the gradient curves coincide within a certain range where [A] > 0.005M. If in fact the order is $1$ or unity, a plot of $[\dot{\mbox{A}}] vs \mbox{[A]}$ would be entirely linear. Fig. (\ref{fig:9}) depict these plots, where some linearity is observed for [A] > 0.005M. 

Fig.(\ref{fig:9}) shows a coincidence of the  rate curves for order  $n=5-7$ above [A]=0.005M. It may be therefore inferred that for at least  $n>5$, and for [A] > 0.005M, the gradient represents the rate constant. The inaccuracy for very low concentrations  may be explained by reference to Fig.(\ref{fig:8}). The experimental curve is  parametrized as 
\begin{equation} \label{eq:14b}
[\mbox{A}]=\sum_{i=1}^{n_r} h_i t^i 
\end{equation}
The time parameter is very large at low concentrations ($>3.5\times 10^5$s) of $\mbox{Bu}^t\mbox{Br}$; the $h_i$ values would be small and the uncertainties in the reactant concentration relatively high; this explains the large scatter in the gradient values at low concentrations. We therefore  ignore the first two  values of [A] at low concentrations and focus on the gradients for points of coincidence  of the different polynomial curves in in  Fig.(\ref{fig:9}). The results are shown in Table \ref{tab:4}. The linear fit in this specified range yields $k_{1d}$. The absolute root mean squared deviation per datapoint of Fig.(\ref{fig:8}) is listed in the last  rhs column of Table \ref{tab:4} where the best fit is in the range $n=5-7$. The average $k_{1d}$ value taking into account the error estimate is $(5.4\pm 0.5)\times 10^{-6}\mbox{s}^{-1}$.
which is close to the $(5.22\pm 0.3)\times 10^{-6}\mbox{s}^{-1}$ of the experimentalists. For the same regime

The results from the literature for this  first order reaction \cite[Table IX,p.2071]{ingold1} using equation (\ref{eq:t1}) has a mean value of $5.22\times 10^{-6} $ and for the 12 datapoints determined , and this value of $k$ varied with range $(5.04-5.48)\times 10^{-6}$ in appropriate dimensions.  The results as computed according our methods in Table \ref{tab:4} 
\begin{table}
	\centering
		\begin{tabular}{|Sl|Sc|Sc|Sl|Sl|}
		\hline
		$n$ & $k_{1d}$/$\mbox{s}^{-1}$ & $k_{1ls}$/$\mbox{s}^{-1}$& Est. error in $k$& Abs. Dev. in $n$ \\ \hline
		2&5.038689$\times 10^{-6}$ & 4.881742$\times 10^{-6}$ & 4.164406$\times 10^{-7}$ & 0.000151 \\ \hline
		3& 5.374684$\times 10^{-6}$ & 5.307609$\times 10^{-6}$ & 8.129860$\times 10^{-8}$ & 0.000073 \\ \hline
		4& 5.053045$\times 10^{-6}$  & 5.047342$\times 10^{-6}$ &3.541633$\times 10^{-7}$ & 0.000034 \\ \hline
		5& 5.398782$\times 10^{-6}$  & 5.181527$\times 10^{-6}$ & 2.772767$\times 10^{-7}$& 0.000014 \\ \hline
		6& 5.393767$\times 10^{-6}$  & 5.184763$\times 10^{-6}$ & 2.723767$\times 10^{-7}$ & 0.000013 \\ \hline
		7& 5.500610$\times 10^{-6}$  & 5.187774$\times 10^{-6}$ & 3.028407$\times 10^{-7}$& 0.000014 \\ \hline
		\end{tabular}
		\caption{Results for the first order $\mbox{Bu}^t\mbox{Br}$ reaction neglecting last datapoint 
		for calculating the mean  rate constant $k$.}	
	\label{tab:4}
\end{table}
 Table \ref{tab:4} shows the absolute deviation per point to be quite small, and Fig.(\ref{fig:8}) shows some plots. The gradient of the curves are graphed for various $n$ in  Fig.(\ref{fig:9}).  The linear fit in this specified range yields $k_{1d}$. The absolute root mean squared deviation per datapoint of Fig.(\ref{fig:8}) is listed in the last  rhs column of Table \ref{tab:4} where the best fit is in the range $n=5-7$. The average $k_{1d}$ value taking into account the error estimate is $(5.4\pm 0.5)\times 10^{-6}\mbox{s}^{-1}$
which is close to the $(5.22\pm 0.3)\times 10^{-6}\mbox{s}^{-1}$ of the experimentalists. For the same regime, (\ref{eq:16}) is used to compute $k_{1ls}$ listed in the 3rd column. For $n=5-7$, $k_{ls}=5.18\times 10^{-6}\mbox{s}^{-1}$ which is exceptionally close to the experimental determination  mentioned above based on the integrated equation (\ref{eq:t1}). requiring initial concentrations. 

\begin{figure}[htbp]
\begin{center}
\includegraphics[width=7cm]{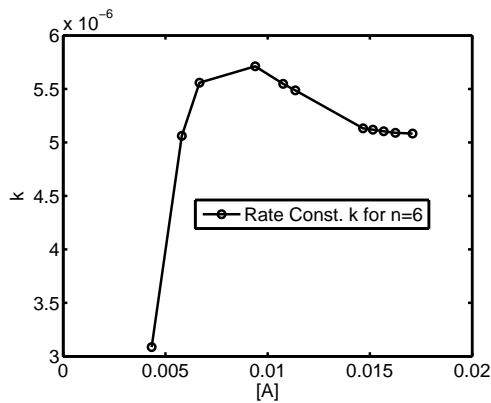} 
\end{center}
\caption{The direct calculation of the change of the rate constant with concentration [A]=$[\mbox{Bu}^t\mbox{Br}]$ directly from the published experimental data \cite[Table IX.]{ingold1} for reaction (iii).}
\label{fig:10} 
\end{figure}

 Lastly, (\ref{eq:15x}) is used to compute the instantaneous rate constant  shown in Fig.(\ref{fig:10}). We note that a maximum is formed before a drop at lower concentrations. Again, the concave form with a maximum is evident here as for reactions (i)-(ii).

\subsection{Reaction (iv) second order details}
\begin{figure}[htbp]
\begin{center}
\includegraphics[width=7cm]{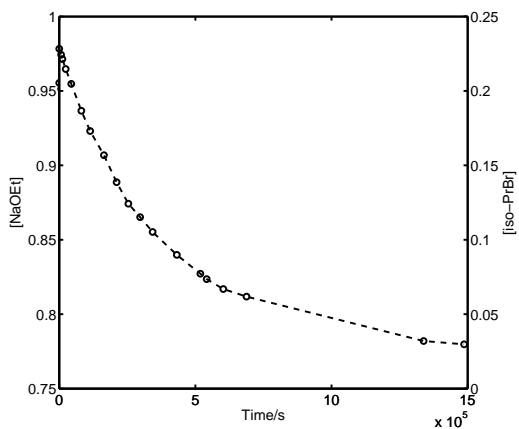} 
\end{center}
\caption{The experimental decay curve for the NaOEt iso-PropylBr $(\mbox{Pr}^i\mbox{Br})$ reaction \cite[Table VII,p.2064]{ingold3} for reaction(iv). }\label{fig:13} 
\end{figure}

\begin{figure}[htbp]
\begin{center}
\includegraphics[width=7cm]{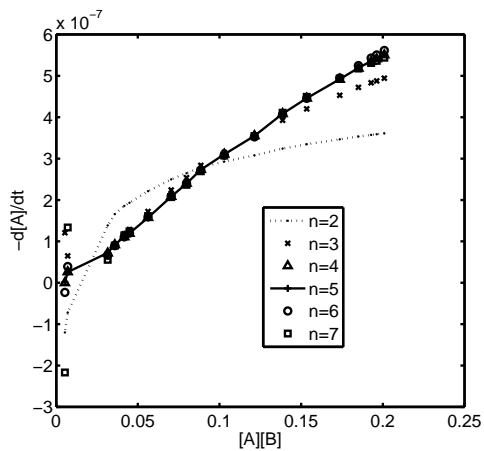} 
\end{center}
\caption{The rate curve shows that the polynomial order n=2-3 is to low to fit  the changes of  gradient. The other curves of higher order all coincide exact for very large time values or low reactant concentrations. The average gradient is calculated by ignoring  these values that are off-scale.$[\mbox{[A]}\equiv\mbox{[NaOEt]}]$ and $[\mbox{[B]}\equiv\mbox{[iso-PropylBr]}]$ in Molar concentration units.  }
\label{fig:11} 
\end{figure}

\begin{figure}[htbp]
\begin{center}
\includegraphics[width=7cm]{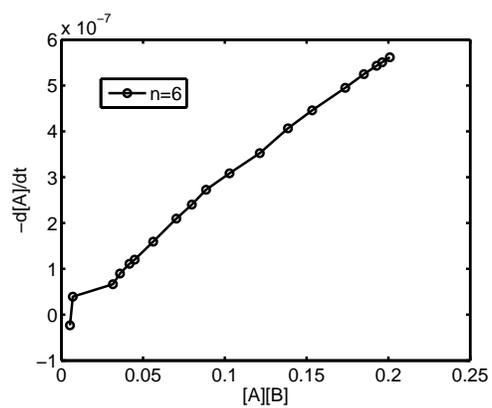} 
\end{center}
\caption{Close-up of the rate curve with reactant product concentration. The last 3 lowest product concentration points are ignored in the rate averaging to determine the rate constant. }
\label{fig:12} 
\end{figure}

\begin{figure}[htbp]
\begin{center}
\includegraphics[width=7cm]{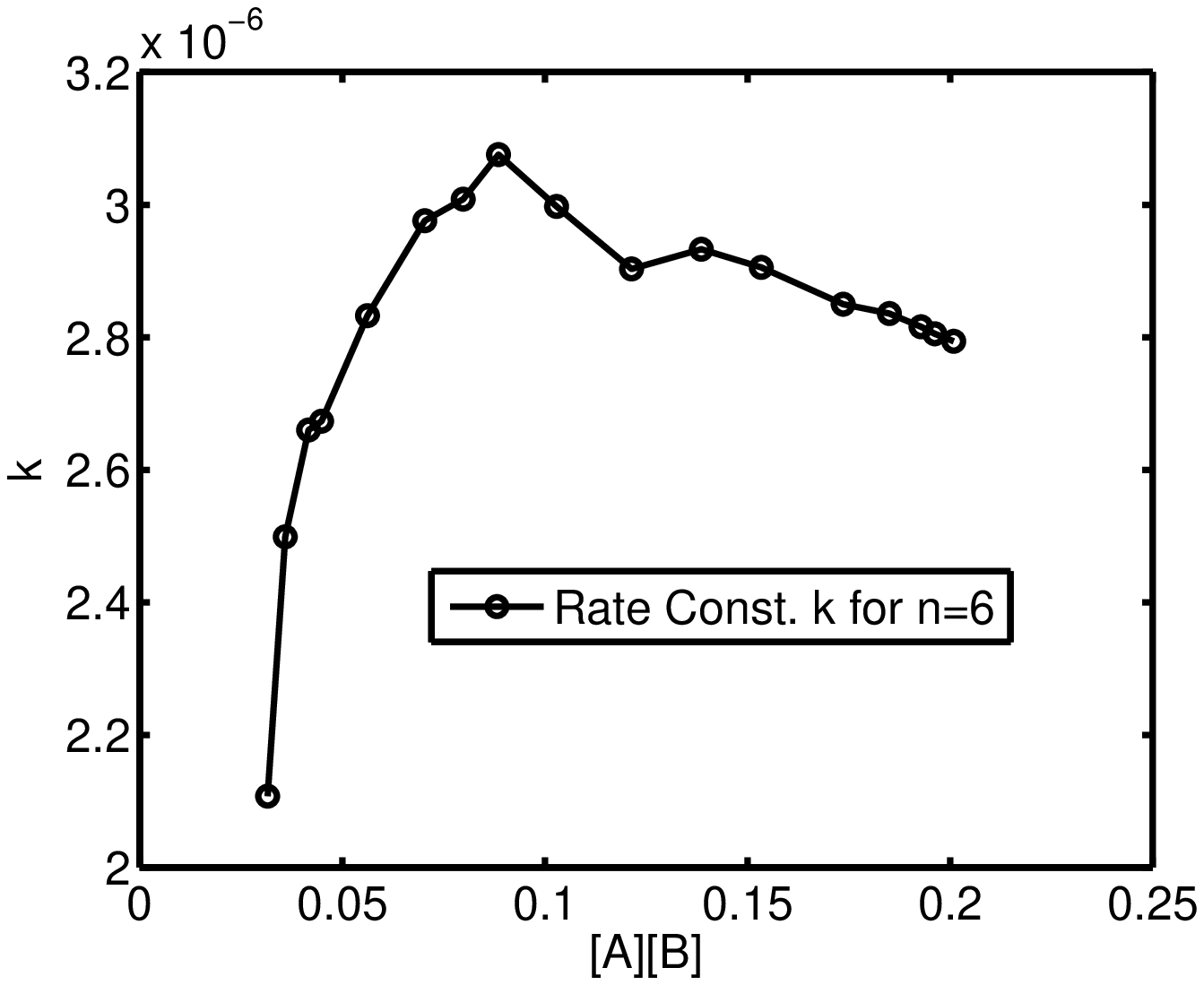} 
\end{center}
\caption{The computation of the instantaneous rate along the regime of coincidence of the polynomials for $n=4-7$ for $n=6$. }
\label{fig:14} 
\end{figure}

The pioneer experimentalists had decided \emph{a priori} that the NaOEt iso-PropylBr $(\mbox{Pr}^i\mbox{Br})$ was second order  and therefore did not independently measure the NaOEt and the $\mbox{Pr}^i\mbox{Br}$ concentrations  as shown in Fig.(\ref{fig:13}). Such a setting introduces  a larger degree of scattering  even if the rate order in known \emph{a priori} to be the appropriate one; the scattering becomes evident in the rate constant curve of Fig.(\ref{fig:14}). The orthogonal polynomial fit  is very good for $n>3$ in Fig.(\ref{fig:13}). The gradient curve of Fig.(\ref{fig:11}) shows a coincidence of points for polynomial order $n>3$ except for the measurement at the lowest concentrations, for the same reasons as given for reaction (iii).  Fig. (\ref{fig:12}) is a close-up of the rate \emph{vs} [A][B] curve where the first 2 points show significant scatter. We ignore the first 3 points of lowest concentrations in our calculations for $k_{2d}$  and $k_{2ls}$ for different polynomial orders n; $k_{2d}$ is the rate constant by linear least squares fit to each of the curves of Fig. (\ref{fig:11}) of d[A]/dt \emph{vs} [A][B] for different polynomial orders $n$  and $k_{2ls}$ is the rate constant calculated according to (\ref{eq:16}). The results are presented in Table \ref{tab:5}. In Table \ref{tab:5}, the value of $k_{2d}$ is remarkably constant for $n=4$ to $7$, in keeping with the curves of Fig. (\ref{fig:11}) that are coincident  for the selected concentration ranges where we have $k_{2d}=(2.80\pm .07)\times 10^{-6}$M$^{-1}$s$^{-1}$ which is very close also to the  computed $k_{2ls}$ values, where the average value  may be written $k_{2ls}=(2.83\pm .05)\times 10^{-6}$M$^{-1}$s$^{-1}$.

\begin{figure}
  \centering
  \subfloat[Reaction (i)]{\label{fig:r1}\includegraphics[width=0.5\textwidth]{m3ratevarytot.eps}}                
  \subfloat[Reaction (ii)]{\label{fig:r2}\includegraphics[width=0.5\textwidth]{n5ratevarykhan.eps}}\\
  \subfloat[Reaction (iii)]{\label{fig:r3}\includegraphics[width=0.5\textwidth]{f9.eps}}
  \subfloat[Reaction (iv)]{\label{fig:r4}\includegraphics[width=0.5\textwidth]{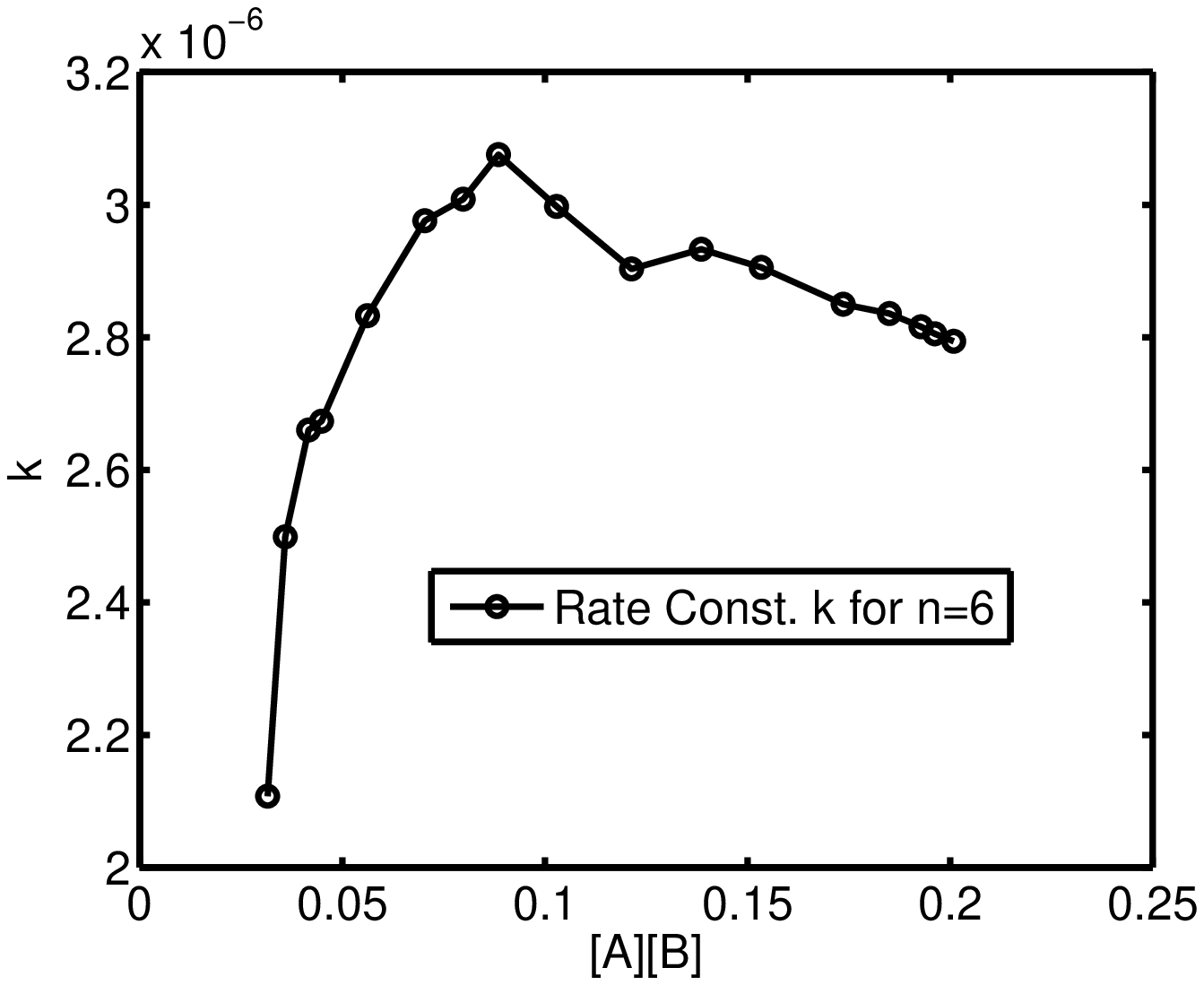}}
  \caption{Collation of  change of rate constants with concentration and time for reactions (i) to (iv).}
  \label{fig:r1to4}
\end{figure}

The experimental results for this second order reaction \cite[Table VII,p.2064]{ingold3} using equation (\ref{eq:t2}) has a mean value of $2.88\times 10^{-6}$M$^{-1}$s$^{-1}$ ($2.95\times 10^{-6}$M$^{-1}$s$^{-1}$ if solvent expansion is taken into account according to an extraneous theory),  and for the 19 datapoints determined ,  $k$ varied within the  range $(2.76-2.96)\times 10^{-6})$M$^{-1}$s$^{-1}$. Fig.(\ref{fig:14}) graphs the instantaneous rate constant for reaction (iv) for for the concentration ranges used for calculating $k_{2d}$ and $k_{2ls}$. Again a semi-sinusoidal shaped  curve is observed. 
There is evident scatter in this graph which may be attributed to the fact that [A] and [B] are completely correlated  and there is no possibility of random  cancellations due to independent measurement of both [A] and [B]. The form of the curve is as for reaction (iii) in Fig.(\ref{fig:10}) \emph{and}  as for reaction (ii) in Fig.(\ref{fig:5b}) if we ignore the low concentration value as having a large scatter at about $t=150$s. Even the ambiguous reaction (i) shows  a shallow concave shape, as with all the rest. Fig.(\ref{fig:r1to4}) is a pallet for the 4  diverse reactions (i)-(iv) of different orders reported by different sources all depicting the same general form; suggestive of some type of overall "chemical inertia"  effect. The  data for   reaction (ii)  were determined over 50 years apart from (iii) and (iv)  by prominent persons, especially Ingold  and co-workers  who were amongst the best kineticists of the 20th century, apart from elucidating  and defining the  SN1, SN2, E1 and E2 reactions in organic chemistry. It seems that all these reactions depict a transient and long-ranging coupling phenomena hitherto unnoticed  due to traditional  analysis that uses integrated rate law expressions with the presupposition of invariant rate constants, which is also built into current statistical mechanics theories. There could be at least two separate possibilities:
 
 (a) independently of the initial concentrations, one observes a rise in the rate constant before falling  when the reactant concentration falls.  Hence if we started a ractin at the midpoint concentrations close to the peak rate constant value in the reaction runs for reactions (ii)-(iv), then the rate constant  profile  will show a form similar to those shown here, except the peak rate constant will shift to lower concentrations, at least  at a concentration less than the concentration at the commencement of the reaction. This possibility suggests some type of "chemical momentum" which is a long-ranging coupling phenomena that current statistical mechanical theories are not able to account for. Obviously this scenario admits the possibility of hysteresis behavior 
 
 (b) the rate constant is simply  a function of reactant and product concentrations as outlined for instance in \cite{cgj1}
  and no hysteresis behavior exists. 

Currently, mainstream statistical mechanical theories do not have a quantum or classical description of (a) and  (b) above. 

Reality would probably be described by broadly either (a) or (b) above with a possibility of the other playing a minor role in the contributed effect.

More careful experiments under stringent conditions need to be performed to :
\begin{description}
\item[$\alpha.$]  verify the existence of this effect by using other methods  just in case the polynomial method has a property that induces a maximum in the gradient at approximately the midpoint of the  domain range under investigation

\item[$\beta.$]  determine which of the two (a)  or (b) above is the preponderant effect.
\end{description}

\begin{table}[htbp]
	\centering
		\begin{tabular}{|Sl|Sc|Sc|Sl|}
		\hline
		$n$ & $k_{2d}$/$\mbox{M}^{-1}\mbox{s}^{-1}$ & Error $\Delta k_{2d}$ & $k_{2ls}/\mbox{M}^{-1}\mbox{s}^{-1}$  \\ \hline
		2& 1.251043$\times 10^{-6}$ & 1.189216$\times 10^{-7}$  & 2.180891$\times 10^{-6}$    \\ \hline
		3&2.142702$\times 10^{-6}$  & 8.845014$\times 10^{-8}$ & 2.654936$\times 10^{-6}$   \\ \hline
		4& 2.818337$\times 10^{-6}$  & 6.850636$\times 10^{-8}$ & 2.829915$\times 10^{-6}$  \\ \hline
		5& 2.822939$\times 10^{-6}$  & 6.600369$\times 10^{-8}$  & 2.833440$\times 10^{-6}$  \\ \hline
		6& 2.876573$\times 10^{-6}$   & 6.613089$\times 10^{-8}$ & 2.853618$\times 10^{-6}$   \\ \hline
		7& 2.770475$\times 10^{-6}$  &7.707235$\times 10^{-8}$  &  2.823186$\times 10^{-6}$ \\ \hline
		\end{tabular}
		\caption{Results for the second order $\mbox{Pr}^i\mbox{Br}$- 
		$\mbox{NaOEt}$ reaction neglecting the last 3 datapoints that are discontinuous  for calculating the   rate constants $k_{2d}$ and $k_{2ls}$.}	
	\label{tab:5}
\end{table}
 
\section{CONCLUSIONS}
The differential methods developed here yield results for the average rate constant that is consistent and close in value to the traditional integrated  law expression for 3  reactions whose rate constants were determined with precision by reliable and prominent kineticists, implying that  the methods developed here are robust. Based on reproducibility of the method in relation to the standard protocol, we further test the differential methods for  the  ambiguous reaction (i) with regard to the inaccurate total reactant concentration (reflected in $\lambda_\infty$) and we showed that the rate constant can be determined  without such data. Since much science refers to systems  whose  key variables are not controllable, as in astrophysical measurements and in forensics and archaeological studies, these methods  could prove useful for analysis in these areas  for reactions to arbitrary order.

All the reactions  studied with different orders and mechanisms  within the polynomial optimization method all show some type of "chemical rate constant momentum" effect in that there is a gradual acceleration in the rate over time initially followed by a sharp decrease as reactants depleted. This observation appears to be novel.  Two independent factors might contribute to this effect:
(A.) the transient coupling of thermodynamical force gradients with flows  and species concentrations  and molecular orientation that leads to the observed profile

(B.) the rate constant is a second order function  of the reactant  and product concentration according to \cite{cgj1}. That theory was based on evidence from \emph{equilibrium} MD simulation of a chemical reaction. 

Stringently controlled   repeated experimentation is required to determine the relative contributions  of (A.) and (B.) above.  If  a semi-sinusoidal profile is observed for different  initial concentrations    that covers the range of the concentration profile  of  a reference reaction of exactly the same type which also exhibits a semi-sinusoidal profile, then (A.) is the predominant mechanism which exhibits hysteresis behavior,  whereas if the resulting  profile  of the experiments leads to a truncated semi-sinusoidal curve  of the reference profile beginning at the initial concentration of the experimental   run, the (B.) is the major effect. There is of course the possibility of a combination of effects (A.) and (B.) to varying degrees. Both effects have not been anticipated in current kinetic theories, which implies that these effects  in themselves  constitutes   one area for further investigation.   

Whilst the main purpose of this work is not to provide physical explanations, we suggest that the resulting curves in Figs.(\ref{fig:r1to4})  can be explained by introducing    X contributing factors or processes in real time. The first (f1) involves the fact that the reactants are initially separated, and the molecules must diffuse  homogeneously  before they can begin to interact. The second process (f2) involves  reactant interaction with the solvent matrix, which would impede the reactive interactions and also conceivably raise the  activation energy relative to unbounded reactants and lastly (f3) describes the product solvent matrix interaction, which is probably is not too significant for the typical reactions  studied here. At the initial stage of the reaction, there is minimal solvent reactant interaction  which would result in the caging of the reaction active sites and so the reaction is diffusion limiting; within a certain time scale, the mutual diffusion of reactants would allow for more reactant-reactant interactions, leading to an  apparent increase in the rate constant; with the progress in time, however, the caging of reactants  due to (f2) would increase the effective activation energy and hence lower the value of the rate constant which explains the precipitous drop at large time values; process (f3) might prevent in some cases the  back reaction due to the breakup of the product to reactant molecules, and it may  moderate (f2) by liming the number of  active solvent interaction with the reactant molecules. One might expect (f1) to case the rise in the rate constant, and (f2) the lowering, leading to the concave maximum observed in all the reactions due to these competing processes. Conventional kinetics, modeled after ideal situations of homogeneity, is not able to account for these fine second  order details in the change in the reaction rate. Even in the homogeneous case , such as a reactive system in thermodynamical equilibrium, it was found that the rate constant is a well-defined function of the reactant and product concentrations, but this contribution to the changes found here is probably of second order for the typical reactions mentioned here.

The results presented here provides alternative developments based on NLA that is able to probe into the finer details of kinetic phenomena than what the standard representations  allow for, especially  in the the areas of changes of the rate constant with the reaction environment as well as determine average rate constant values without $\lambda_\infty $being known. Even with the assumption of invariance of $k$, one can always choose the best type of polynomial order that is consistent with the assumption, and it appears that the initial concentration  as well as the rate constant seems be be predicted as global properties based on the polynomial expansion.It should be noted that the examples chosen here were first order ones; the methods  are general and they pertain to any form of rate law where the gradients and forms can be curve-fitted and the form of the equations optimized as in section (\ref{ss2}). One other research area that may be investigated is the possibility of reactions of fractional order; elementary reactions are by nature of integer order; is there a method that can reduce them to fractional order if  the rate constant is indeed  in part a weak  function of the reactant concentrations?

\section{ACKNOWLEDGMENTS}
This work was supported by University of Malaya Grant UMRG(RG077/09AFR) and Malaysian Government grant   FRGS(FP084/2010A).  \\ \\

\renewcommand{\refname}{BIBLIOGRAPHY}
 \bibliographystyle{unsrt}	

\bibliography{mpbib}
\end{document}